\documentclass[prb,twocolumn]{revtex4}
\usepackage{graphicx}
\usepackage{epsfig}
\begin{document}

%\draft

%\begin{multicols}{2}

\title{Dualism in Entanglement and Testing Quantum to Classical Transition of Identicity}
\author{S. Bose$^1$ and D. Home$^2$}

\affiliation{$^1$Department of Physics and Astronomy, University
College London, Gower St., London WC1E 6BT, UK}
\affiliation{$^2$Physics Department, Bose Institute, Calcutta
700009, India}

\begin{abstract}
\end{abstract}

\maketitle

%\begin{multicols}{2}
Identical objects in quantum mechanics, which are
indistinguishable through their static attributes (mass, charge
etc.), are known to differ from their classical counterparts by
exhibiting quantum statistics
\cite{hanbury,hong,zeilinger97,yamamoto98}. Here we show a
hitherto unexplored consequence of the property of identicity in
quantum mechanics. If two identical objects, distinguished by a
dynamical variable ${\cal A}$, are in certain entangled states of
another dynamical variable ${\cal B}$, then, for such states, they
are also entangled in variable ${\cal A}$ when distinguished from
each other by variable ${\cal B}$. This dualism is independent of
quantum statistics. Departures from identicity of the objects due
to arbitrarily small differences in their innate attributes
destroy this dualism. A system independent scheme to test the
dualism is formulated which is readily realizable with photons.
This scheme can be performed without requiring the entangled
objects to be brought together. Thus whether two macro-systems
behave as quantum identical objects can be probed without the
complications of scattering. Such a study would complement the
program of testing the validity of quantum superposition principle
in the macro-domain \cite{macro1,macro2,macro3,macro4} which has
stimulated considerable experimentation
 \cite{macro5,macro6,macro7,macro8,macro9,macro10}.

   The concept of identicity of
objects in quantum mechanics is characterized by the sameness of
the magnitudes of their static attributes, such as mass, charge,
or magnetic moment. Thus identical quantum particles can be
distinguished from each other {\em only} through their dynamical
variables. For example, given two electrons, unless they differ in
either their spin, position, momentum, energy or some such
dynamical variable, they are indistinguishable. Whenever we refer
to two identical objects as object-1 and object-2, the labels 1
and 2 actually stem from distinctness of the objects in some
dynamical variable.  A typical example is the EPR-Bohm entangled
state of two spin-$1/2$ particles \cite{bohm}
\begin{equation}
|\Psi\rangle_{12}=\frac{1}{\sqrt{2}}(|\uparrow\rangle_1|\downarrow\rangle_2+|\downarrow\rangle_1|\uparrow\rangle_2)
\label{bohm}.
\end{equation}
In the above, if the particles are identical, the labels $1$ and
$2$ cannot correspond to any innate difference (such as mass or
charge) between them, and have to correspond to their different
position or momentum states. So given an entangled state of two
identical particles, to meaningfully describe it, we need at least
two variables: One variable ${\cal A}$ to distinguish the
particles, and another variable ${\cal B}$ which is entangled.

        Let us now
consider, using precisely the same notation as in Eq.(\ref{bohm}),
the following entangled state of two identical particles
\begin{equation}
|\Psi\rangle_{A_1A_2}=\alpha~|B_1\rangle_{A_1}|B_2\rangle_{A_2}+
\beta~ |B_2\rangle_{A_1}|B_1\rangle_{A_2}, \label{state1}
\end{equation}
where $A_1$ and $A_2$ are {\em distinct} eigenvalues of variable
${\cal A}$, $B_1$ and $B_2$ are {\em distinct} eigenvalues of
variable ${\cal B}$ and $\alpha$ and $\beta$ are non-zero complex
amplitudes. In the above, ${\cal A}$ acts as the dynamical
variable (or group of commuting variables) which serves as the
identity label of the particles (such as position) and ${\cal B}$
acts as the variable which is entangled (such as spin). As the
particles are identical, $|\Psi\rangle_{A_1A_2}$ can be rewritten
in the second quantized notation as
\begin{equation}
|\Psi\rangle_{A_1A_2}=(\alpha~
c^\dagger_{A_1,B_1}c^\dagger_{A_2,B_2}+\beta~
c^\dagger_{A_1,B_2}c^\dagger_{A_2,B_1})|0\rangle, \label{second}
\end{equation}
where $|0\rangle$ is the vacuum state for the particles under
consideration, and $c^\dagger_{A_i,B_j}$ signifies the creation of
a particle in the eigenstate $|A_i,B_j\rangle$ of the variables
${\cal A}$ and ${\cal B}$. We now carry out the following simple
manipulation
\begin{eqnarray}
|\Psi\rangle_{A_1A_2}&=&(\alpha~
c^\dagger_{A_1,B_1}c^\dagger_{A_2,B_2}\pm
\beta~c^\dagger_{A_2,B_1}c^\dagger_{A_1,B_2})|0\rangle \label{step1}\\
&=&\alpha~|A_1\rangle_{B_1}|A_2\rangle_{B_2}\pm \beta~
|A_2\rangle_{B1}|A_1\rangle_{B2}, \label{step2}
\end{eqnarray}
where in Eq.(\ref{step1}), the order of the last two creation
operators have been interchanged with appropriate signs (the upper
sign stands for bosons and the lower for fermions), and in
Eq.(\ref{step2}), ${\cal B}$ is chosen to be the variable which
serves as the particle identity label.

    The two representations of the state $|\Psi\rangle_{A_1A_2}$ given
by Eqs.(\ref{state1}) and (\ref{step2}) bring out a {\em dualism
in entanglement for identical particles}. If two identical
particles, distinguished by a dynamical variable ${\cal A}$, are
entangled in another dynamical variable ${\cal B}$ in a way as
described in Eq.(\ref{state1}), then they are also entangled in
variable ${\cal A}$ when distinguished from each other by variable
${\cal B}$, as evident from Eq.(\ref{step2}). Eq.(\ref{step2})
also implies that the {\em magnitude} of the dual of entanglement
for the state under consideration is {\em same} for both bosons
and fermions (the magnitude in this case can be quantified by the
degree of violation of a Bell's inequality). Thus the above
dualism is a more {\em general} property of identical particles
than their specific quantum statistics.

  An alternative derivation of the
dualism can be made using first quantized notation with {\em
pseudo labels} $i,j$ for the particles to
symmetrize/antisymmetrize their joint state. The entangled state
$|\Psi\rangle_{A_1A_2}$ of Eq.(\ref{state1}) can be rewritten as
\begin{eqnarray}
|\Psi\rangle_{A_1A_2}&=&\alpha~(|A_1,B_1\rangle_i|A_2,B_2\rangle_j\pm
|A_2,B_2\rangle_i|A_1,B_1\rangle_j)\nonumber\\&+&\beta~(|A_1,B_2\rangle_i|A_2,B_1\rangle_j\pm
|A_2,B_1\rangle_i|A_1,B_2\rangle_j)\nonumber\\
&=&|A_1\rangle_i|A_2\rangle_j(\alpha~|B_1\rangle_i|B_2\rangle_j+\beta~|B_2\rangle_i|B_1\rangle_j)\nonumber\\
&\pm &
|A_2\rangle_i|A_1\rangle_j(\alpha~|B_2\rangle_i|B_1\rangle_j+\beta~|B_1\rangle_i|B_2\rangle_j)\label{pseudo1}\\
&=&(\alpha~|A_1\rangle_i|A_2\rangle_j\pm
\beta~|A_2\rangle_i|A_1\rangle_j)|B_1\rangle_i|B_2\rangle_j
\nonumber\\
&+&(\alpha~|A_2\rangle_i|A_1\rangle_j\pm
\beta~|A_1\rangle_i|A_2\rangle_j)|B_2\rangle_i|B_1\rangle_j.\label{pseudo2}
\end{eqnarray}
Eq.(\ref{pseudo1}) tells that if $A_1$ and $A_2$ are used as {\em
true labels} of the $i$th and the $j$th (or the $j$th and the
$i$th) particles respectively, then they are in the entangled
state given by Eq.(\ref{state1}). Eq.(\ref{pseudo2}) implies its
dual with the roles of ${\cal A}$ and ${\cal B}$ interchanged.
There is a {\em complementarity} in the sense that we {\em cannot
observe both the entanglements at the same time}. One has to use
either the variable ${\cal A}$ or the variable ${\cal B}$ as a
``which-particle" label and the entanglement in the other variable
can then be observed.

\begin{figure}
\begin{center}
\leavevmode \epsfxsize=3in \epsfbox{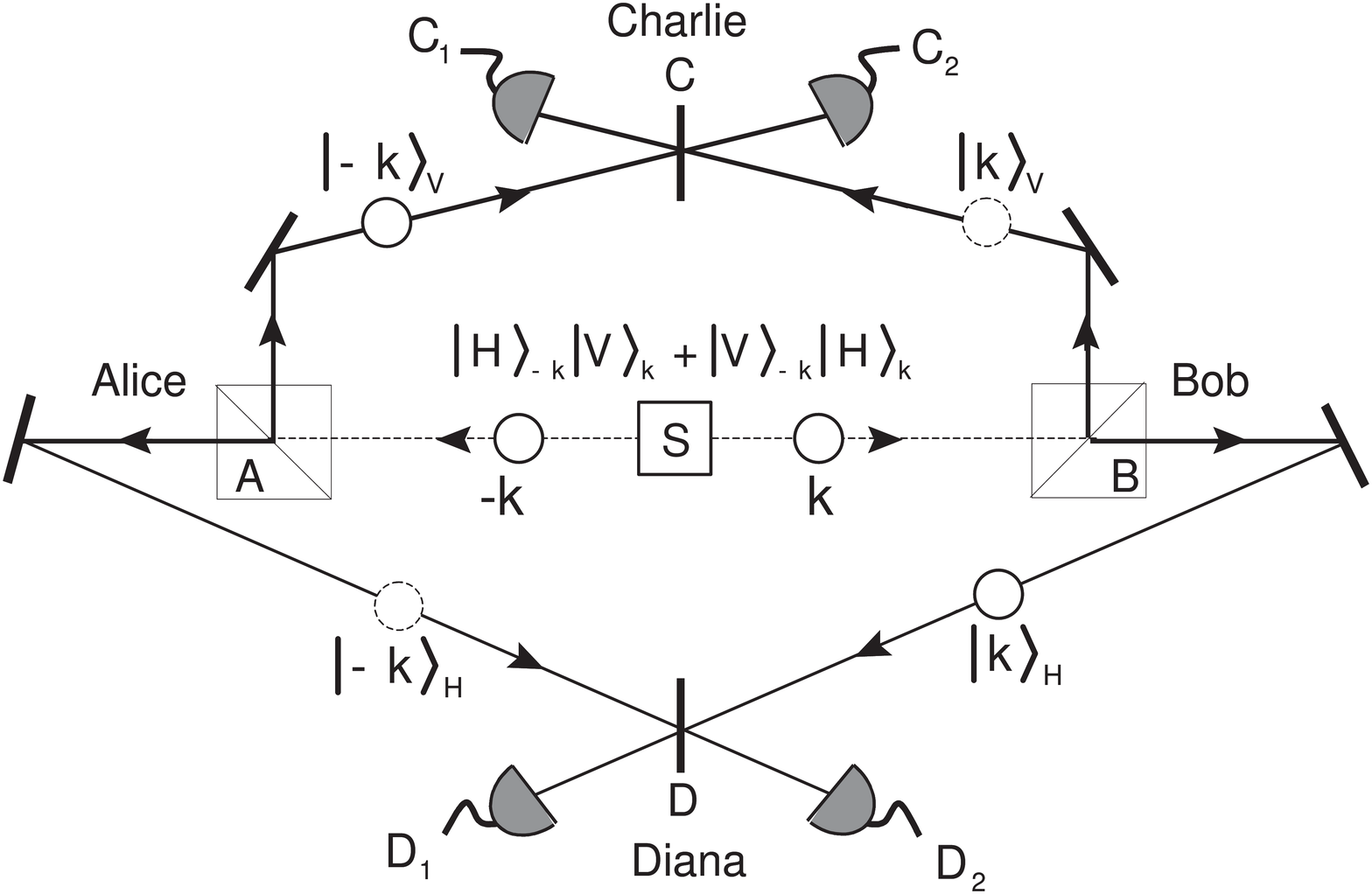}
 \caption{Scheme to test the dualism in entanglement of identical particles.
 Two photons of the same frequency in the polarization entangled state $|\Psi\rangle_{12}=\frac{1}{\sqrt{2}}(|H\rangle_{1}|V\rangle_{2}+|V\rangle_{1}|H\rangle_{2})$ are emitted in opposite directions by the
source S. The labels $1$ and $2$  stand for their momenta {\bf -k}
and  {\bf k} respectively, along the $x$ axis. The {\bf -k} and
{\bf k} photons fly towards
 Alice and Bob respectively. If they conducted
 polarization correlation measurements on these photons they could verify their polarization
 entanglement. Instead, in order to verify the {\em dual} momentum entanglement implied by the right hand side
 of Eq.(\ref{dualexp}), they put polarization
 beam splitters A and B in the paths of the photons which deflects
 $|V\rangle$ photons
 in the $+y$ direction  (towards Charlie), and $|H\rangle$ photons in the $-y$ direction (towards Diana). Because the
 photon pair is emitted in the state $|\Psi\rangle_{12}$, only one photon reaches Charlie and one photon reaches Diana
 per pair and by the dualism of Eq.(\ref{dualexp}), the $x$-component of their momenta is
entangled. Charlie and Diana can now conduct measurements on the
photons reaching them and verify the entanglement.
 As there are only two possible values of
 the
 $x$-component of the momentum, namely {\bf -k} and {\bf k}, one can associate a dichotomic pseudo-spin
 observable \cite{home,hasegawa} with this momentum. These pseudo-spin operators and their linear combinations
 are
 measurable
 by a beam splitter with a tunable reflectivity and two detectors. Charlie and Diana are
 each equipped with
 precisely such a momentum pseudo-spin measuring apparatus. By noting
 coincidence of clicks in their detectors, Charlie and
 Diana can verify whether a Bell's inequality is violated by
 momentum pseudo-spin correlation measurements on their photons and
 thereby test our dualism.
 }
\label{dual1}
\end{center}
\end{figure}

    Note that for the rather {\em special case} $\alpha=\beta$, the
state $|\Psi\rangle_{A_1A_2}$ assumes the factorizable form
$(|A_1\rangle_i|A_2\rangle_j\pm
|A_2\rangle_i|A_1\rangle_j)(|B_1\rangle_i|B_2\rangle_j+|B_2\rangle_i|B_1\rangle_j)$,
(similar result holds for $\alpha=-\beta$). This is known and
exploited for Bell state measurements in quantum teleportation
\cite{bouwmeester}. Such experiments, however, do not probe the
dualism in entanglement (interchangeability of ${\cal A}$ and
${\cal B}$). Moreover, this dualism is more general than the above
factorizability in the sense that it persists even for $\alpha\neq
\pm\beta$.

 Some entangled states of identical
particles will, however, not {\em manifestly} exhibit the dualism.
For example, the state $\alpha~|B_1\rangle_{A_1}|B_1\rangle_{A_2}+
\beta~ |B_2\rangle_{A_1}|B_2\rangle_{A_2}$ will not exhibit
dualism as the particles concerned are not in distinct eigenstates
of the variable ${\cal B}$. However, a local rotation
($|B_1\rangle_{A_2}\leftrightarrow |B_2\rangle_{A_2}$) can convert
the state to one that exhibits dualism.  We also stress that if
the entangled objects were non-identical (due to, say, an {\em
arbitrarily small} mass difference), one would have to replace the
right hand side of Eq.(\ref{second}) by $(\alpha~
c^\dagger_{A_1,B_1}d^\dagger_{A_2,B_2}+\beta~
c^\dagger_{A_1,B_2}d^\dagger_{A_2,B_1})|0\rangle|0\rangle$ where
$c^\dagger$ and $d^\dagger$ create particles from different
vacuua. Our dualism will {\em not} hold any more. Testing the
dualism is thus {\em equivalent} to establishing that such objects
behave as identical quantum objects ({\em i.e.,} are created from
the {\em same} vacuum).

\begin{figure}
\begin{center}
\leavevmode \epsfxsize=3in \epsfbox{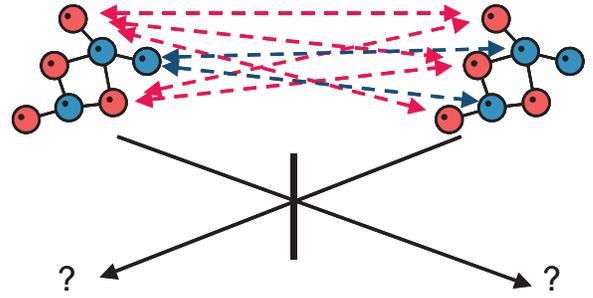}
 \caption{The difficulty of testing whether two complex
 objects (such as molecules) can behave as identical objects in the quantum mechanical sense. Existing methods \cite{hanbury,hong,zeilinger97,yamamoto98,bose-home} involve interfering two
 such objects at a beam-splitter to observe their quantum statistical
 behavior. However, when two such objects are prepared in identical states and brought
 together, they may interact and undergo a complicated scattering with amplitudes in all directions, the result of
 which is extremely difficult to calculate. Moreover,
 the two complex objects may undergo an inelastic collision,
 split into fragments and/or undergo a chemical reaction. Thus,
 without making them non-interacting, one cannot test their
 ability to behave as quantum identical objects using the existing
 schemes.
 }
\label{dual2}
\end{center}
\end{figure}

     The above dualism can be tested with polarization
entangled photons in the state $|\Psi\rangle_{12}$ with $1$ and
$2$ corresponding to momenta labels ${-\bf k}$ and ${\bf k}$
respectively, and $\uparrow$ and $\downarrow$ standing for
orthogonal polarization states $H$ and $V$ respectively. The
dualism can then be expressed as
\begin{equation}
|H\rangle_{\bf -k}|V\rangle_{\bf k}+|V\rangle_{\bf
-k}|H\rangle_{\bf k}\equiv |{\bf -k}\rangle_{H}|{\bf
k}\rangle_{V}+ |{\bf k}\rangle_{H}|{\bf -k}\rangle_{V}.
\label{dualexp}
\end{equation}
Let ${\bf k}$ be chosen to be along the $x-$axis. Then the
polarization entanglement implied by the left hand side of
Eq.(\ref{dualexp}) can be tested in the usual manner by Alice and
Bob on opposite locations along the $x-$axis. For testing its
dual, we separate the $H$ and $V$ components of the state along
the $y-$axis. Then Charlie and Diana on opposite locations along
the $y-$axis can verify the momentum entanglement. The details are
given in Fig.\ref{dual1} and its caption. The same test is
possible with any pair of entangled identical objects (atoms,
electrons, neutrons or even macro-molecules), with two states of
some internal degree of freedom replacing $H$ and $V$ in
Eq.(\ref{dualexp}). A practical application of the dualism
stemming from the above scheme is that it enables the {\em usage}
of a spin/polarization entangled state, if it is easier to produce
for certain classes of particles, as a momentum entangled state.

   In all known procedures for testing the identicity of quantum objects
(such as bunching and anti-bunching), the particles need to be
brought together. However, in testing our dualism according to the
above scheme, the objects in the $|{\bf k}\rangle$ and the $|{\bf
-k}\rangle$ state are {\em never} present together at a specific
location. Thus our scheme for testing the dualism offers a way of
testing the identicity of two quantum objects without bringing
them together. The caveat, of course, is that they have to be in
an entangled state {\em before} the experiment. If we use
mediating particles to entangle the objects {\em prior} to the
experiment, then we can always keep them well separated. For
example, one can test the ability of two distant photons to behave
as quantum identical objects without bringing them together by
first entangling them using entanglement swapping \cite{entswp},
and then using our scheme.

\begin{figure}
\begin{center}
\leavevmode \epsfxsize=3in \epsfbox{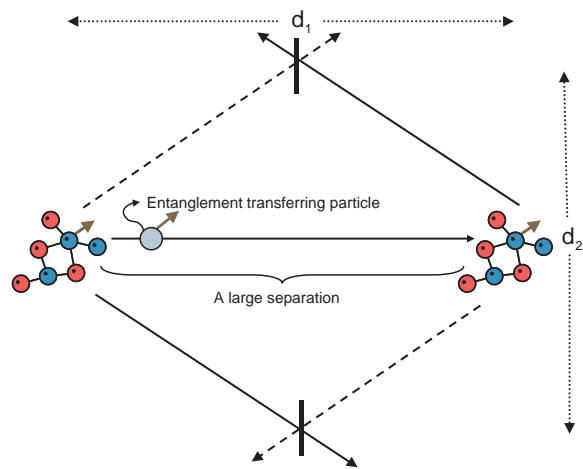}
 \caption{Setup to probe the quantum to classical transition of
 identicity using the dualism in entanglement. Two identical macro-molecules are cooled in well separated
traps and prepared in
 a maximally
 entangled state of some internal dynamical variable such as internal vibrations or rotations or the spin on
 one of the constituent atoms. Two orthogonal states of this variable replace $|H\rangle$ and $|V\rangle$ in Eq.(\ref{dualexp}). The entanglement between the molecules can be
 mediated by a particle which interacts successively with
 a specific
 variable of the two molecules as shown, and need not involve bringing the molecules together. The molecules are then ejected from their
 traps in the state of Eq.(\ref{dualexp}) and scheme described in
 Fig.\ref{dual1} is performed to test their identicity through the
 dualism of
 entanglement. Under this scheme, the pair of molecules take a superposition of the dotted and the bold
 paths as shown and have no amplitude to come close and
 interact. As the molecules used in the experiment are made larger, we expect a gradual/sudden loss of the ability of the molecules to behave as identical particles in the quantum mechanical
 sense (whatever the mechanism). This will be
manifested as a lowering/vanishing of the violation of the Bell's
inequality testing the dual entanglement in the momentum degrees
of freedom. We expect this effect to be enhanced by simply keeping
the molecules separated longer so that they interact with distinct
environments. If this is done by lengthening the time before the
molecules go into the superposition of paths then the effect can
be made to depend more strongly on $d_1$ relative to $d_2$. Any
decoherence
 of the superposition of the paths of the molecules will also reduce the degree of violation
 of the same Bell's inequality, but this will depend more symmetrically on $d_1$ and $d_2$.}
\label{dual3}
\end{center}
\end{figure}

   The above procedure will become particularly relevant for testing the quantum identicity of two complex objects
(such as two macro-molecules), since it avoids the complications
due to scattering which rule out the existing tests
\cite{hanbury,hong,zeilinger97,yamamoto98,bose-home}, as explained
in Fig.\ref{dual2} and its caption. Our scheme adapted to
macro-molecules is shown in Fig.\ref{dual3}. After verifying the
dualism with two identical molecules, one could perform our scheme
with two molecules of the same species differing in mass by a {\em
tiny} amount (such as two
 $C_{60}$ isotopes \cite{macro9}) and verify that the
dualism is destroyed. Of course, as in {\em any scheme} to test
quantum identicity, the molecules should be cooled to a
temperature $\theta \leq \hbar^2/2m k_B (\Delta x)^2$, where
$\hbar$ and $k_B$ are Planck's and Boltzmann's contants
respectively, $m$ is the molecular mass, and $\Delta x$ is the
position spread of the wavefunction of each of the molecules, so
that they are not distinguishable through their arrival time at
the detectors (for a molecule of mass number $100$ cooled in a nm
sized trap $\theta\sim 1$mK).

The transition from the quantum to the classical world through
decohering superpositions is a well studied issue both
theoretically \cite{caldiera,zurek} and experimentally
\cite{macro8,wineland}. In contrast, the transition from what we
mean by identical particles in the quantum world to what we mean
by identical particles in the classical world remains unexplored.
It is true that the superposition principle is involved in the
symmetrization procedure of quantum mechanics, but this form of
superposition cannot be destroyed by the {\em usual}
system-environment interaction of decoherence models
\cite{caldiera,zurek} as the superposed components (say, $|{\bf k}
\rangle_i|{\bf -k}\rangle_j$ and $|{\bf -k} \rangle_j|{\bf
k}\rangle_i$) differ only in {\em pseudo} labels $i$ and $j$ and
  do not induce distinct evolutions of the environment. It is, however,
possible that two mesoscopic identical objects will interact with
{\em distinct}
   environments when kept well separated. They will thereby evolve to distinct (orthogonal)
states of
   variables other than the ones involved in any experiment to test their quantum identicity. This will
   effectively be a
quantum to classical transition
  of identicity which can be studied systematically through our scheme as explained in the caption of Fig.\ref{dual3}.

      The connection of our dualism with
recent work on entanglement quantification in identical particle
systems could be of future interest
\cite{schliemann,zanardi,omar,fisher,vaccaro,vedral}. Modelling
the quantum to classical transition of identicity theoretically by
coupling two identical particles in identical states to distinct
environments, should be an important future program.

\textbf{\small Acknowledgements.} {\small We thank J. I. Cirac, J.
Preskill, Y. Yamamoto, W. E. Lawrence, C. Simon, M.-A. Dupertuis
and G. Adesso for insightful remarks after various talks by SB on
the subject matter of the letter. DH acknowledges the Jawaharlal
Nehru Fellowship and the Royal Society-Indian National Science
Academy Exchange Programme which made this collaboration
possible.}

\vspace{1cm}
%\end{multicols}

%

\end{document}